\documentclass{PoS}
\usepackage{graphicx}
\usepackage{amsmath}
\usepackage{epsfig}
\usepackage{slashed}

\title{QCD calculations with optical lattices? }

\ShortTitle{QCD calculations with optical lattices? }

\author{\speaker{Y. Meurice}\\ 

Department of Physics and Astronomy, The University of Iowa\\
Iowa City, Iowa 52242, USA\\
         E-mail: \email{yannick-meurice@uiowa.edu}\\}
\abstract{By trapping cold polarizable atoms in periodic potentials created by crossed laser beams, it is now possible to experimentally create "clean" lattice systems. Experimentalists have successfully engineered local and nearest-neighbor interactions that approximately recreate Hubbard-like models on table tops. I discuss the possibility of using this new technology in the context of lattice gauge theory and in particular, relativistic dispersion relations, flavor symmetry, functional derivatives and emerging local gauge symmetry.}

\FullConference{ The XXIX International Symposium on Lattice Field Theory - Lattice 2011\\
July 10-16, 2011\\
Squaw Valley, Lake Tahoe, California}

\begin{document}
\section{Introduction}
Recently, experimental atomic physicists have been able to trap cold polarizable atoms in periodic potentials created by crossed laser beams. They can  create "clean" lattice systems and have successfully engineered local and nearest-neighbor interactions that approximately recreate Hubbard-like models, often used in condensed matter, on table tops.  I discuss the possibility of using the same technology 
for lattice gauge theory. I first review the basic features of optical lattices. For a recent review article, I recommend  Ref. \cite{RevModPhys.80.885}. I then discuss some of the basics tools for model building (literally!). For a more comprehensive ``toolbox" see Ref. \cite{2005AnPhy.315...52J}. 
Atomic physics is a rapidly expanding field where theorists and experimentalists work closely together \cite{2010NatPh...6..998T}. Up to now, the possibility of creating dynamical gauge fields is quite open \cite{2011arXiv1101.5369M}. I will make some suggestions at the end. 
\section{Optical lattices}
Alkali-metals (Li, Na, K, Rb, Cs) are often used in cold atoms experiments because of their loosely bound electron in the the outer shell. Typical choices are  $^{87}Rb$ (a boson: 37 e$^-$, 37 p and 50 n) or $^{6}Li$ (a fermion: 3 e$^-$, 3 p and 3 n). The polarizable cold atoms are trapped in 
standing waves created by counterpropagating laser beams in 1, 2 or 3 dimensions. The periodic potential is due to the dipole moment induced by the linearly polarized laser beam and reads: 
\begin{equation} 
V({\bf r})=
-(1/2)\alpha(\omega) |{\bf E}({\bf r})|^2\ ,
\end{equation}
with 
\begin{equation}
\alpha(\omega)  \sim |<e|d|g>|^2/\hbar(\omega_0-\omega_L) \  .
\end{equation} 
For instance, it is possible to create a 3D lattice potential with a cubic symmetry by using 3 mutually orthogonal laser beams of the same  wavelength $\lambda_L$. 
The periodic potential reads
\begin{equation}
\label{eq:per}
V(x,y,z)=V_0(\sin ^2(kx)+\sin^2 (ky)+\sin ^2 (kx)) \  ,
\end{equation}
 with $k=2\pi/\lambda_L$. The lattice spacing is $a=\lambda_L/2$ . The depth of the potential $V_0$ 
is measured in units of the recoil energy $E_r\equiv (\hbar k)^2/2 m _{\rm atom}$ and can be tuned continuously by changing the intensity of the laser. 

In order to fix the ideas,  we consider the experimental realization of the Bose-Hubbard model of Ref. \cite{2002Natur.415...39G}.  
Rubidium atoms were used with a laser operating at a wavelength $\lambda_L =856$nm. The recoil energy is $1.3 \times 10^{-11} \  {\rm eV} \simeq k_B 1. 5 \times  10^{-7}$K. 
For reference, the critical temperature for Bose condensation in Rubidium with a specific volume of $(\lambda_L/2)^3$
 is close to $10^{-7}$K according to the ideal gas formula. The recoil momentum  is $1.5$eV/c and the recoil velocity about 5 mm/s. 
Of  the order of $N_{atoms} \simeq 65^3$ were used. Assuming one atom per site, the physical size of the lattice is of the order of 30 $\mu m$. 
It takes a few milliseconds for atoms moving at speeds of the order of the recoil velocity to go across the lattice size. 
The depth of the potential $V_0$ was increased up to 20 $E_r$. For this maximal value, the harmonic frequency is approximately 30 kHz. 

The Bose-Hubbard model can be used  to describe a set of bosonic atoms trapped in the potential (\ref{eq:per}). 
The Hamiltonian  reads
  \begin{equation}
H = -J\sum_{\langle i,j \rangle}( a^{\dagger}_{i} a^{}_{j}+ h.c.) + U \sum_{i=1} n_{i} (n_{i}-1) \   .
\end{equation}
The notation ${\langle i,j \rangle}$ means the sum over nearest neighbors on a cubic lattice. The $a_i$ and their conjugate obey bosonic commutation relations, and $n_i=a^{\dagger}_ia_i$. 
The phase diagram of this model with a chemical potential is discussed in Ref. \cite{PhysRevB.40.546}.
The constant $J$ and $U$ can be expressed in terms of integrals involving the Wannier functions for the periodic potential (\ref{eq:per}).
Explicit expressions can be found in section II of Ref. \cite{2010NatPh...6..998T}. For a shallow enough potential, $J>>U$ and tunneling 
dominates, leading to a superfluid phase. On the other hand, for $V_0$ large enough, $J<<U$,  tunneling is suppressed and we are in the Mott insulator phase. 
For more details see Ref. \cite{RevModPhys.80.885}.

  \section{What can we learn from experiments with optical lattices?}
  
  One important experimental method to learn about the spacial correlations functions 
  \begin{equation}
  G({\bf x-y})=<a^{\dagger}_{\bf x}a_{\bf y}> \  ,
  \end{equation}
   is to suddenly release the optical potential. 
  If the interactions after the release can be neglected, the state of the system will evolve according to its plane wave decomposition. After a given time of flight $t$, the density distribution $n({\bf x})$ 
  can be imaged and interpreted according to the relation
  \begin{equation}
  n(x)=\left( \frac{M}{\hbar t}\right )^3|\hat{w}({\bf k})|^2 \hat{G}({\bf k}) \ ,
  \end{equation}
with ${\bf k}=M{\bf x}/(\hbar t)$. In the superfluid phase (small enough $V_0/E_r$), the phase coherence produces sharp peaks at reciprocal lattice vectors. 
As $V_0$ is increased, the interference peaks surrounding the zero momentum peak become more pronounced. Across the quantum phase transition to the Mott insulator phase, this trend gets reversed, but the interference peaks subside for a while. 
Deep in the Mott insulator phase ($V_0>>E_r$),  the distribution becomes Gaussian. 

 At finite temperature and unity filling, the phase diagram of the Bose-Hubbard model in the $(U/J,T)$ plan may look vaguely reminiscent to the QCD one 
 in the $(\mu, T)$ plane (as far as the shape is concerned). 
 A schematic picture of the phase diagram, a discussion of the experimental setup and theoretical calculations as well as appropriate references are provided in Ref. \cite{2010NatPh...6..998T}. 
 In the homogeneous case, the superfluid phase is bounded by an approximate quarter of circle surrounding (0,0). In practice, the inhomogeneities due to the trapping 
 potential may play an important role. Experimentally, the temperature is raised by controlled heating sequences up to temperatures of 400 nK. Theoretically, the Bose-Hubbard Hamiltonian with 
 a site-dependent chemical potential can be simulated with the worm algorithm. The experimental and simulated time of flight distributions for various values of $U/J$ and $T$ are shown in Fig. 3 of Ref. \cite{2010NatPh...6..998T} and 
 show good agreement between theory and experiment. 

 \section{Model building}
 
The example discussed in the previous section shows that it is possible to build a physical system that approximates a lattice many-body Hamiltonian.   
Can we modify this setup in order to get several flavors of fermions with relativistic dispersion relations? 
Using three of the 6 hyperfine levels from the  F=1/2 and 3/2 states of a $^6$Li Fermi gas near a Feshbach resonance, one 
can create a quantum degenerate three-state Fermi gas with approximate $SU(3)$ symmetry \cite{williams-2009}.  These three states leads to interesting scattering length patterns for magnetic fields between 500 and 1000 Gauss. 

Relativistic dispersion relations for the fermions can be obtained by using ``bicolored'' lattices. 
After introducing Fourier modes, the dispersion relation is obtained by solving a quadratic equation. The fact that two possible signs for the square root of the 
discriminant leads to Dirac cones. 
One well-known example is graphene \cite{2011RSPTA.369.2625V}. It is also possible to meet such a requirement by creating different potential depths on alternate sites \cite{PhysRevLett.105.190403}. Interesting ways of coupling Dirac fermions to periodic or staggered gauge potentials by combining two types of square lattices have also been 
proposed in Refs. \cite{PhysRevA.82.013616, PhysRevA.81.033622}. More generally, many interesting tools are available for theoretical and experimental ``model building". For a ``toolbox" see  Ref.  \cite {2005AnPhy.315...52J}.

\section{Dynamical Gauge fields?}
 
We now discuss the most important ingredient: gauge fields. 
Background gauge fields can be generated by rotating the optical lattice. In a rotating frame, we have the ``minimal substitution" 
\begin{equation}
\frac{d \vec{x}}{dt} \rightarrow  \frac{d \vec{x}}{dt} -\vec {\Omega} \times \vec{x} \  ,
\end{equation}
 which corresponds to a constant magnetic field $e\vec{B}= 2m \vec{\Omega}$. 
 Global non-abelian Berry phases can be obtained from adiabatic transformations in degenerate quantum mechanical systems 
\cite{PhysRevLett.52.2111}.  Such phases can be obtained from ``dark states" in a tripod system
\cite{juzeliunas-2008-100}. Global $SU(N)$ potentials can also be created using $N$ internal states of atoms and laser assisted state sensitive tunnelling
\cite{PhysRevLett.95.010403}. All these constructions are global, however, locally rotating deformations of optical lattice 
have been studied recently \cite{gemelke-2010}. 

As the essential physical properties at short and large distances come from gauge fields, it seems essential  to be able to design experimental setups with dynamical gauge fields. 
An idea that would come naturally to many particle physicists is to build the link variable $U_{{\bf x},i}\ ^{ab}$ as a ``condensate" of the site variables ${\phi}_{{\bf x}}^a\ $ at the ends of the link \cite{2011arXiv1101.5369M}:
\begin{equation}
U_{{\bf x},{\bf e}_i}^{ab}={{\phi}^{\star}}_{\bf x}^a \phi_{{\bf x+ e}_i} ^b \  .
\end{equation}
After searching the literature, we found models that have a chance to be implemented on optical lattices where such a situation occurs  
\cite{PhysRevB.37.627,PhysRevB.38.745,PhysRevB.38.2926}. The starting point is the (Fermi-)Hubbard model with Hamiltonian
\begin{equation}
\label{eq:hub} 
H = -t \sum_{\langle i,j \rangle,\alpha}( f^{\dagger}_{i,\alpha} f^{}_{j,\alpha}+ h.c.) + U \sum_{i=1}^{N} n_{i\uparrow} n_{i\downarrow} \ ,
\end{equation}
where $t$ characterizes tunneling to nearest neighbor sites and $U>0$ controls the onsite repulsion. The $f^{}_{j,\alpha}$ and $f^{\dagger}_{i,\alpha}$ satisfy 
anticommutation relations and $\alpha$ is a spin index that takes two values. 

We now restrict ourselves to half-filling (number of fermions = number of sites) and two space dimensions. 
At $t=0$, and for $N$ sites, there are $2^N$ ground states  with exactly one fermion at every site since neighbor spins have no influence on each other.
In the limit $U>>t$,  
the cost of a single tunnelling is small, but the cost of double occupancy is large. At second order in $t$, it is possible to exchange two neighbor fermions 
without affecting the single occupancy constraint. Using a canonical transformation to eliminate the first term in Eq. (\ref{eq:hub}) and neglecting terms of order higher than $t^2$, one obtains, up to a constant, the Heisenberg model with $J=4t^2/U$ (this $J$ should not be confused with the one used for the Bose-Hubbard model ):
\begin{equation}
H = J\sum_{<{i}{j}>} \mathbf{S}_{i}\cdot \mathbf{S}_{j} \  ,
\end{equation}
with
\begin{equation}
 \mathbf{S}_{i} = \frac{1}{2} f_{{i}\alpha}^\dagger \mathbf{\sigma}_{\alpha \beta} f_{{i}\beta} \  ,
\end{equation}
\noindent
or equivalently 
\begin{equation}
\nonumber
H=\sum_{<{i}{j}>}-\frac{1}{2} J f_{{i}\alpha}^\dagger f_{{j}\alpha} f_{{j}\beta}^\dagger f_{{i}\beta} +\sum_{<{i}{j}>} J (\frac{1}{2} n_{i}-\frac{1}{4} n_{i} n_{j})
\label{eq:h}
\end{equation}
\noindent
with the constraint
\begin{equation}
f_{{i}\alpha}^\dagger f_{{i}\alpha} =1 \   .
\label{eq:n}
\end{equation}
This model has a $SU(2)$ space-dependent local gauge-invariance. This property and other interesting features are discussed in Refs.  \cite{PhysRevB.37.627,PhysRevB.38.745,PhysRevB.38.2926,wen}. 
The connection with the Hamiltonian formulation of LGT (one  doublet of fermions coupled to $SU(2)$ gauge fields) in the temporal gauge can be  made more clear by 
rewriting 
the four fermion interactions in $H$ using auxiliary fields attached to the links $(i,j)$ and coupled linearly to 
fermions bilinear (one at each end of the link). 
Using the Lagrangian formulation (with a continuous time), it is possible to introduce Lagrange multipliers that enforce the condition (\ref{eq:h}), play the role 
of time-like $SU(2)$ gauge fields and make the Lagrangian invariant under time-dependent gauge transformations.  

Note that the correspondence between the original Hubbard model and the LGT model is only valid at the lowest nontrivial order in the strong coupling expansions (large $U$ for Hubbard 
at half-filling, large $g$ for LGT). In particular, standard plaquette or other interactions could appear at higher order. Finding many-body Hamiltonians that can be experimentally implemented and which are equivalent to LGT models as we know them is a challenge  for the future. A possibility suggested by Cheng Chin is to use two lattices one having 
molecules that can hop and induce the desired interactions on the other lattice. 
Quantum link formulations \cite{Chandrasekharan:1996ih} might also suggest possible implementations of optical lattice models with local gauge invariance. 

\section{Functional derivatives?}

In modern quantum field theory, the $n$-point functions are obtained as functional derivatives.  Generically, we can write 
\begin{equation}
<ABC\dots >=\frac{\partial}{\partial J_A}\frac{\partial}{\partial J_B}\frac{\partial}{\partial J_C}\dots  \ln Z(\{ J\})|_{\{ J\}=0}\  ,
\end{equation}
for an Hamiltonian $H_J=H+AJ_A+BJ_B+CJ_C \dots $. Very recently, single atom manipulations have been performed experimentally
\cite{2011Natur.471..319W}. Changes in free energy due to atom removal might provide information regarding correlation functions. 

\section{Conclusions}
In conclusion, the study of optical lattices is a new and exciting area of interest to all lattice practitioners.
The experimental implementation of  standard QFT tools (functional derivatives ...) needs to be developed. 
The question of dynamical gauge fields is wide open and could lead to an interesting model-building effort involving several lattice communities. 

\vskip10pt
\noindent
{\bf Acknowledments}

I thank Gordon Baym, Cheng Chin, Nathan Gemelke, Yuzhi Liu, Cristiane Morais-Smith, Ken O' Hara, Nikolay Prokofiev, Boris Svitsunov, Eddy Timmermans and Shan-Wen Tsai for conversations and suggestions. 



\begin{thebibliography}{10}

\bibitem{RevModPhys.80.885}
I.~Bloch, J.~Dalibard, and W.~Zwerger, ``Many-body physics with ultracold
  gases,'' {\em Rev. Mod. Phys.} ~80, ~885--964, Jul 2008.

\bibitem{2005AnPhy.315...52J}
D.~{Jaksch} and P.~{Zoller}, ``{The cold atom Hubbard toolbox},'' {\em Annals
  of Physics} ~315, ~52--79, Jan. 2005.

\bibitem{2010NatPh...6..998T}
S.~{Trotzky}, L.~{Pollet}, F.~{Gerbier}, U.~{Schnorrberger}, I.~{Bloch}, N.~V.
  {Prokof'Ev}, B.~{Svistunov}, and M.~{Troyer}, ``{Suppression of the critical
  temperature for superfluidity near the Mott transition},'' {\em Nature
  Physics} ~6, ~998--1004, Dec. 2010.

\bibitem{2011arXiv1101.5369M}
Y.~{Meurice}, ``{Dynamical Gauge Fields on Optical Lattices: A Lattice Gauge
  Theorist Point of View},''  {arXiv.org:1101.5369}, Jan. 2011.

\bibitem{2002Natur.415...39G}
M.~{Greiner}, O.~{Mandel}, T.~{Esslinger}, T.~W. {H{\"a}nsch}, and I.~{Bloch},
  ``{Quantum phase transition from a superfluid to a Mott insulator in a gas of
  ultracold atoms},'' {\em Nature} ~415, ~39--44, Jan. 2002.

\bibitem{PhysRevB.40.546}
M.~P.~A. Fisher, P.~B. Weichman, G.~Grinstein, and D.~S. Fisher, ``Boson
  localization and the superfluid-insulator transition,'' {\em Phys. Rev. B}
  ~40, ~546--570, Jul 1989.

\bibitem{williams-2009}
J.~R. Williams, E.~L. Hazlett, J.~H. Huckans, R.~W. Stites, Y.~Zhang, and K.~M.
  O'Hara, ``Evidence for ground- and excited-state efimov trimers in a
  three-state fermi gas,'' 2009.
\newblock arXiv.org:0908.0789.

\bibitem{2011RSPTA.369.2625V}
M.~A.~H. {Vozmediano}, ``{Renormalization group aspects of graphene},'' {\em
  Royal Society of London Philosophical Transactions Series A}~369,
  ~2625--2642, July 2011.

\bibitem{PhysRevLett.105.190403}
J.~I. Cirac, P.~Maraner, and J.~K. Pachos, ``Cold atom simulation of
  interacting relativistic quantum field theories,'' {\em Phys. Rev. Lett.}
  ~105, p.~190403, Nov 2010.

\bibitem{PhysRevA.82.013616}
L.-K. Lim, A.~Lazarides, A.~Hemmerich, and C.~Morais~Smith, ``Competing pairing
  states for ultracold fermions in optical lattices with an artificial
  staggered magnetic field,'' {\em Phys. Rev. A} ~82, p.~013616, Jul 2010.

\bibitem{PhysRevA.81.033622}
X.-J. Liu, X.~Liu, C.~Wu, and J.~Sinova, ``Quantum anomalous hall effect with
  cold atoms trapped in a square lattice,'' {\em Phys. Rev. A} ~81,
  p.~033622, Mar 2010.

\bibitem{PhysRevLett.52.2111}
F.~Wilczek and A.~Zee, ``Appearance of gauge structure in simple dynamical
  systems,'' {\em Phys. Rev. Lett.} ~52, ~2111--2114, Jun 1984.

\bibitem{juzeliunas-2008-100}
G.~Juzeliunas, J.~Ruseckas, A.~Jacob, L.~Santos, and P.~Ohberg, ``Double and
  negative reflection of cold atoms in non-abelian gauge potentials,'' {\em
  Physical Review Letters} ~100, ~200405, 2008.

\bibitem{PhysRevLett.95.010403}
K.~Osterloh, M.~Baig, L.~Santos, P.~Zoller, and M.~Lewenstein, ``Cold atoms in
  non-abelian gauge potentials: From the hofstadter "moth" to lattice gauge
  theory,'' {\em Phys. Rev. Lett.}  ~95, ~010403, Jun 2005.

\bibitem{gemelke-2010}
N.~Gemelke, E.~Sarajlic, and S.~Chu, ``Rotating few-body atomic systems in the
  fractional quantum hall regime,'' 2010.
\newblock arXiv.org:1007.2677.

\bibitem{PhysRevB.37.627}
Z.~Zou and P.~W. Anderson, ``Neutral fermion, charge-$e$ boson excitations in
  the resonating-valence-bond state and superconductivity in
  ${\mathrm{la}}_{2}$${\mathrm{cuo}}_{4}$-based compounds,'' {\em Phys. Rev.
  B} ~37, ~627--630, Jan 1988.

\bibitem{PhysRevB.38.745}
I.~Affleck, Z.~Zou, T.~Hsu, and P.~W. Anderson, ``Su(2) gauge symmetry of the
  large-$u$ limit of the hubbard model,'' {\em Phys. Rev. B} ~38,
  ~745--747, Jul 1988.

\bibitem{PhysRevB.38.2926}
E.~Dagotto, E.~Fradkin, and A.~Moreo, ``Su(2) gauge invariance and order
  parameters in strongly coupled electronic systems,'' {\em Phys. Rev. B}
  ~38, ~2926--2929, Aug 1988.
  
\bibitem{wen}
  X.-G. Wen, ``Quantum Filed Theory of Many-Body Systems", Oxford University Press, 2004.

\bibitem{Chandrasekharan:1996ih}
S.~Chandrasekharan and U.~J. Wiese, ``{Quantum link models: A discrete approach
  to gauge theories},'' {\em Nucl. Phys.} ~B492, ~455--474, 1997.

\bibitem{2011Natur.471..319W}
C.~{Weitenberg}, M.~{Endres}, J.~F. {Sherson}, M.~{Cheneau}, P.~{Schau{\ss}},
  T.~{Fukuhara}, I.~{Bloch}, and S.~{Kuhr}, ``{Single-spin addressing in an
  atomic Mott insulator},'' {\em Nature} ~471, ~319--324, Mar. 2011.

\end{thebibliography}
\end{document}